\documentclass[twocolumn,aps,prb,reprint,showpacs,amsmath,amssymb]{revtex4-1}
\usepackage{epsfig}
\usepackage{graphicx}
\usepackage{dcolumn}
\usepackage{bm}
\usepackage{multirow}
\usepackage{CJK}
\usepackage{float}

\begin{document}
\begin{CJK*}{GBK}{}
\title{Nonequilibrium critical dynamics in the quantum chiral clock model}

\author{Rui-Zhen Huang$^1$}
\email{huangrzh@foxmail.com}
\author{Shuai Yin$^2$}
\email{zsuyinshuai@163.com}
\affiliation{$^1$Kavli Institute for Theoretical Sciences, University of Chinese Academy of Sciences, Beijing 100190, China}
\affiliation{$^2$Institute for Advanced Study, Tsinghua University, Beijing 100084, China}
\date{\today}

\begin{abstract}
In this paper we study the driven critical dynamics in the three-state quantum chiral clock model. This is motivated by a recent experiment, which verified the Kibble-Zurek mechanism and the finite-time scaling in a reconfigurable one-dimensional array of $^{87}$Rb atoms with programmable interactions. This experimental model shares the same universality class with the quantum chiral clock model and has been shown to possess a nontrivial non-integer dynamic exponent $z$. Besides the case of changing the transverse field as realized in the experiment, we also consider the driven dynamics under changing the longitudinal field. For both cases, we verify the finite-time scaling for a non-integer dynamic exponent $z$. Furthermore, we determine the critical exponents $\beta$ and $\delta$ numerically for the first time. We also investigate the dynamic scaling behavior including the thermal effects, which are inevitably involved in experiments. From a nonequilibrium dynamic point of view, our results strongly support that there is a direct continuous phase transition between the ordered phase and the disordered phase. Also, we show that the method based on the finite-time scaling theory provides a promising approach to determine the critical point and critical properties.
\end{abstract}

\maketitle

\section{\label{intro}Introduction}
One of the most fascinating arenas in modern theoretical physics and condensed matter physics is to understand the non-equilibrium evolution in quantum many-body systems~\cite{Dz,Pol,Rig}. Recent studies on the Rydberg-atomic systems shed new light on this vibrant field~\cite{Lukin}. Lots of fascinating dynamic properties have been discovered from various frameworks~\cite{Lukin}. For the relaxation dynamics, it has been shown that these systems can escape the usual thermal fate~\cite{Dun}, which is described by the celebrated eigenstate thermalization hypothesis~\cite{Deu,Sre,Rig1}, due to the appearance of the embedded quantum scarred states~\cite{Tur,Tur1,Khe,Lin,Choi}. For the driven dynamics~\cite{Keesling}, the Kibble-Zurek mechanism (KZM)~\cite{Kibble,Zurek} and the finite-time scaling (FTS) theory~\cite{Zhong1,Zhong2,Yin1} are verified in a one-dimensional ($1$D) reconfigurable array of $^{87}$Rb atoms experimentally. In addition to the quantum Ising universality class, this experiment also bears witness to the validation of the KZM and the FTS in the $\mathbb{Z}_3$ quantum chiral clock model (QCCM)~\cite{Huse,Ostlund,Fendley}, which can be obtained by mapping the translational symmetry of the Rydberg-atomic system into the internal symmetry of the clock system~\cite{Fendley,Sachdevchi1,Sachdevchi}.

These sustained efforts, devoted into the Rydberg-atomic systems and the QCCM, are expected to be quite worthwhile, since the QCCM has potential applications in the quantum information technology. A kind of exotic excitation---the \textit{parafermion} mode, which is often regarded as a sister of the Majorana zero mode, is hosted in the QCCM~\cite{Fendley1,Fendley2,Mong,Alicea,Mazza,Calzona}. Similar to the Majorana mode, this parafermion mode is also considered as a promising candidate for the topological quantum computation~\cite{Mong,Alicea}.

\begin{figure}[tbp]
\includegraphics[angle=0,scale=0.23]{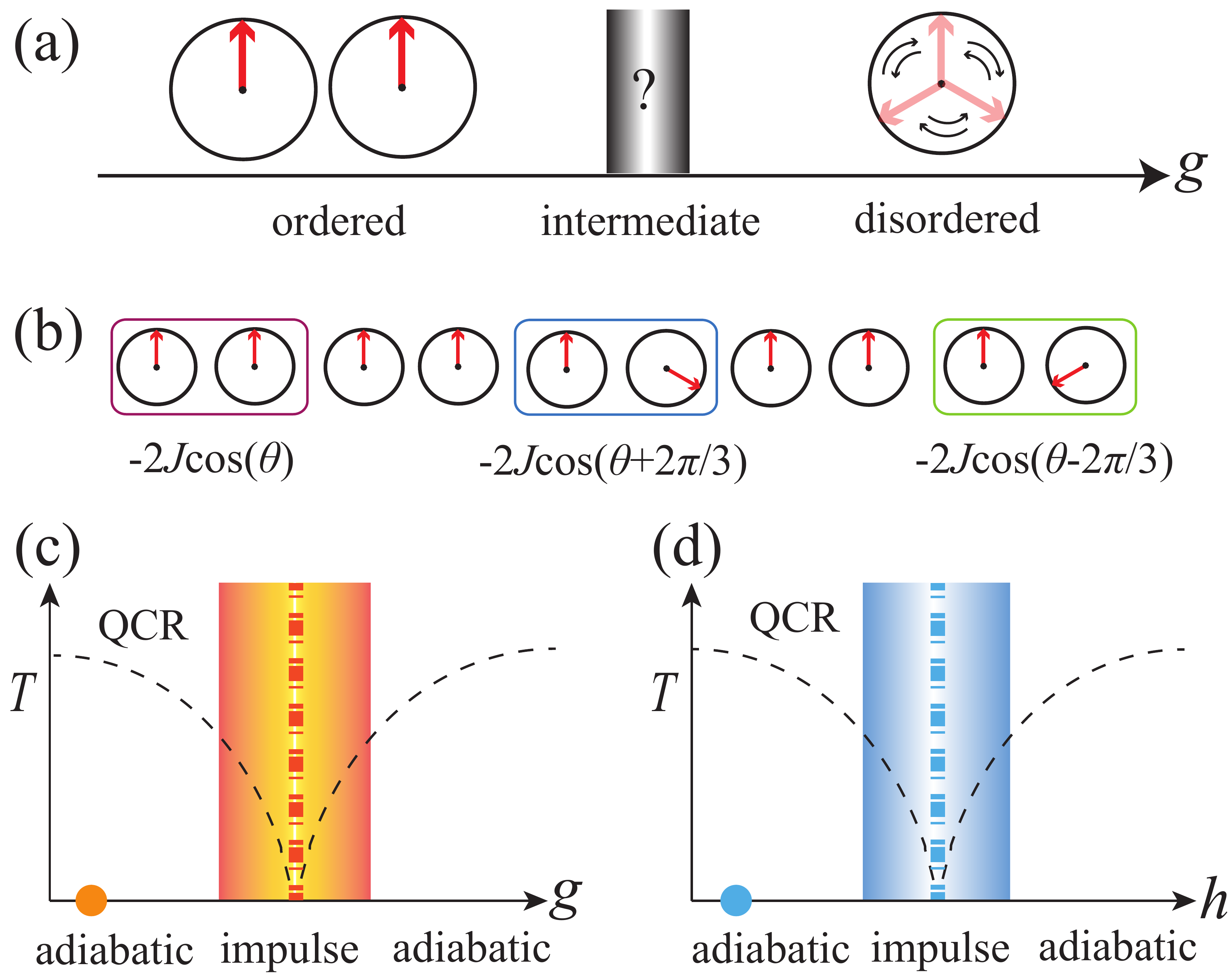}
  \caption{Schematic figure of the model and protocols of the driven dynamics. (a) The phase diagram with a dubious intermediate region between the ordered and disordered phases. (b) The schematic representation of the three kinds of bond interactions in the QCCM. (c) The protocol of the driven dynamics by changing the transverse field $g$. (d) The protocol of the driven dynamics by changing the transverse field $h$. In addition to the case of the conventional KZM, which requires an adiabatic initial stage (orange and blue dots on the abscissa axis), this paper also considers the driven dynamics starting from a thermal equilibrium state (orange and blue columns) near the critical point. The QCR marks the quantum critical region.}
\label{fig:clock}
\end{figure}

Besides, the QCCM has its own theoretical interests since it possesses a rich phase diagram with appealing phase transition behaviors. A long-standing puzzle, as illustrated in Fig. \ref{fig:clock}, is whether there exists an intermediate phase between the ordered and disordered phases in the $1$D $\mathbb{Z}_3$ QCCM. A direct phase transition was proposed in an early work~\cite{Huse,Huse1}, but was subsequently questioned~\cite{Haldane}. Until recently, the state-of-the-art numerical works provide the vindication for a direct phase transition~\cite{Zhuang,Samajdar,Chepiga}. This conclusion is then supported by a field-theoretic renormalization group study~\cite{Sachdevchi}. A remarkable feature of this critical point is that its dynamic exponent $z$, obtained from both numerical~\cite{Samajdar} and analytic works~\cite{Sachdevchi}, is a non-integer value, implying an underlying nonconformal critical theory. Moreover, this property of $z$ is then confirmed experimentally by the FTS collapse of correlation functions with different driving rates~\cite{Keesling}. While very common in classical critical dynamics~\cite{Tauber}, a non-integer dynamic exponent $z$ is highly nontrivial in the quantum case~\cite{Sachdev,Sondhi}, since the static and dynamic properties are intertwined therein via the imaginary-time path integral. It is natural to ask whether the conventional dynamic scaling theory, for instance, the KZM or the FTS, is still applicable. Inspired by the experimental progress~\cite{Keesling}, systematic studies on the driven quantum critical dynamics in the QCCM are hence called for.

In this paper, we consider various protocols of the driven critical dynamics in the $1$D $\mathbb{Z}_3$ QCCM (See Fig.~\ref{fig:clock}) with time-reversal symmetry. In addition to the case of changing the transverse field, which is realized in the experiment~\cite{Keesling}, we also consider the driven dynamics of changing the longitudinal field. For both cases, we confirm that the quantum FTS is applicable with a nontrivial $z$. As the classical counterpart of the QCCM lives in a nonexistent noninterger space dimension, critical exponents $\beta$ and $\delta$, to the best of our knowledge, have not been determined before~\cite{Huse}, neither in recent numerical studies of the QCCM~\cite{Samajdar}. By applying the FTS theory, we complete the table of the critical exponents by estimating all critical exponents, including $\beta$ and $\delta$, numerically. From this nonequilibrium aspect, our results consolidate the conclusion of the direct continuous phase transition between the ordered and disordered phases. Furthermore, considering that the thermal effects are inevitably involved in real experiments, we investigate the driven dynamics staring from a thermal equilibrium state near the critical point, as illustrated in Fig.~\ref{fig:clock}. A modified FTS, which includes the initial conditions as its additional scaling variables, is then confirmed numerically.

The rest of this paper is organized as follows. The quantum chiral clock model and the numerical method are introduced in Sec.~\ref{sec_model}. Then a brief review of the KZM and the FTS is given in Sec.~\ref{KZMFTS}. In Sec.~\ref{sec_t_0}, we show our results and compare them with previous studies. In Sec.~\ref{sec_t}, we further study the effects induced by the thermal initial condition. Finally a summary is given in Sec.~\ref{sec_summary}.

\section{Model and numerical method}\label{sec_model}
\subsection{$\mathbb{Z}_3$ quantum chiral clock model}
The Hamiltonian of the $\mathbb{Z}_n$ QCCM in one dimension (See Fig.~\ref{fig:clock}) reads
\begin{equation}
    \mathcal{H} = -f \sum_i \tau_i^\dagger e^{-i\phi} - J \sum_i \sigma_i^\dagger \sigma_{i+1} e^{-i\theta} + \rm{h.c.},
    \label{Eq_model}
\end{equation}
in which $f>0$ and $J>0$, and $\sigma$ dictates the direction of the watch hand, while $\tau$ rotates the watch hand clockwise through a discrete angle $2\pi/n$. $\sigma$ and $\tau$ satisfy the algebra $\sigma_i^n=\mathbb{I}$, $\tau_i^n=\mathbb{I}$, and $\sigma_i \tau_j = \omega \delta_{ij} \tau_j \sigma_i$, where $\omega = e^{2\pi i/n}$. A global $\mathbb{Z}_n$ transformation represented by $\mathcal{G}\equiv \prod_i \tau_i$ makes the Hamiltonian invariant.

In the full parameter space, the subspace of $n=3$ has attracted special attentions~\cite{Zhuang,Samajdar,Chepiga}. In the basis of $\sigma$, the explicit matrices for $\sigma$ and $\tau$ are
\begin{equation}
    \sigma =
\left(
    \begin{array}{ccc}
        1  &  0      &  0\\
        0  & \omega  &  0\\
        0  &  0      &  \omega^*
    \end{array}
\right)
, \tau =
\left(
    \begin{array}{ccc}
        0  &  1  &  0\\
        0  &  0  &  1\\
        1  &  0  &  0
    \end{array}
\right).
\end{equation}
For $\theta=\phi=0$, this model reduces to the quantum Potts model~\cite{Potts}. The phase transition from the $\mathbb{Z}_3$-broken phase to the symmetric phase is described by the $c=4/5$ conformal field theory. The driven dynamics in this model has been studied and the KZM and the FTS are verified therein~\cite{qijun,Ghosh}. In this paper, we will focus on the QCCM with $\theta\neq0$ and $\phi=0$. For $f\ll J$, the system is in an ordered phase, which breaks the $\mathbb{Z}_3$ symmetry, while for $f\gg J$ the system is in a disordered phase. However, controversy appears in the region in between. For large $\theta$, it is an incommensurate phase. This phase shrinks as $\theta$ decreases. However, it is not clear whether this incommensurate phase penetrates into the vicinity of zero $\theta$. Until recently, studies based on the high-precision numerical method show that this incommensurate phase ceases to evolve at a finite $\theta$~\cite{Zhuang,Samajdar,Chepiga}. Hence, for small $\theta$, a direct phase transition happens from the disordered phase to the ordered phase. This has been further supported by a renormalization group theory~\cite{Sachdevchi}. Furthermore, both numerical and analytic results demonstrate that the dynamic exponent $z$ is a non-integer value~\cite{Zhuang,Samajdar,Chepiga,Sachdevchi}. And this nontrivial property of $z$ is also found in the experiment~\cite{Keesling}.

In the following, we will mainly consider the case for $\phi=0$ and $\theta=\pi/8$. Previous studies~\cite{Zhuang,Samajdar} show that these parameters give a direct continuous phase transition between the ordered phase and the disordered phase. We also set $J=1$ and adopt it as unit of energy. The distance to the critical point $g$ is defined as $g\equiv f-f_c$ with $f_c$ being the phase transition point, at which the order parameter $m\equiv\langle \sigma_i + \sigma_i^\dagger \rangle$/2 vanishes as $m\sim (-g)^\beta$ in equilibrium. As shown in Fig~\ref{fig:clock}, we also consider the case including in model~(\ref{Eq_model}) an additional symmetry-breaking term $-h\sum_i \sigma_i + \rm{h.c.}$ with $h$ being a longitudinal field. At $g=0$, $m$ satisfies $m\sim h^{1/\delta}$.

\subsection{Numerical method and calculation setup} \label{sec_method}
The numerical methods utilized in this paper are based on the infinite matrix product state (MPS)~\cite{Cirac,Schollwock}, which decompose the full quantum state into the multiplication of matrices. In this way, each site is attached by a set of matrices. By optimizing these matrices, one can obtain a well-approximated state with limited costs. It has been demonstrated that the MPS method is a quite efficient tool in studying the static~\cite{Cirac2} and dynamic behaviors~\cite{Schollwock2} for $1$D quantum systems.

For the zero temperature simulation, we first need to obtain the ground state. To do this, we use the variational uniform MPS approach~\cite{ref_vmps}. It has been demonstrated that the MPS with a given dimension $D$ of its matrix at each site forms a submanifold of the whole Hilbert space~\cite{ref_diag_transfer_mat}. Thus one can efficiently obtain the converged ground state wave function by variationally optimize the matrix of an MPS through minimizing the ground state energy within this submanifold. Then for the time evolution, we use the infinite time-evolving block decimation method~\cite{ref_i_tebd}. The matrix of the MPS is updated by imposing the local evolution operator, which is the Suzuki-Trotter decomposition of the full time evolution operator $\textrm{exp}(-i\mathcal{H}t)$.

For the finite temperature simulation, we first need to obtain the thermal equilibrium state by the purification technique\cite{ref_puri_mps}. An auxiliary system maximally entangled with the physical system is introduced in the MPS and the density operator can be explicitly restored by tracing out the auxiliary system. Starting from infinite temperature represented by an MPS, by doing imaginary time evolution one can obtain the thermal state $ |\Psi(2T)\rangle$.  Then one can carry real time evolution by acting time evolution operators on $|\Psi(2T)\rangle$ and obtain the MPS $|\Psi(2T,t)\rangle$ at any time $t$. The expectation value for a physical quantity $\hat{O}$ at time $t$ and temperature $T$ is then calculated according to $\langle \hat{O} \rangle = \langle \Psi(2T,t) | \hat{O} | \Psi(2T,t) \rangle$, where $| \Psi(2T,t) \rangle $ is normalized already.

To ensure the accuracy of the numerical simulation, we make the following setup for the MPS calculation. In evaluating the ground state wave function, the gradient\cite{ref_vmps} of the ground state energy with respect to the matrix in the MPS is required to be smaller than $10^{-14}J$. During the time evolution process at zero and finite temperatures, the time interval is taken to be $1\times 10^{-2}$. And the fourth order Suzuki-Trott decomposition is used, which ensures the Trotter error in one single time step is about $10^{-10}$. $D$ of the MPS used in our calculation is generally $100$. $D$ up to $300$ is checked and no apparent correction is observed. It has been demonstrated that during the whole driven process, the correlation length and entanglement entropy of the system is always finite (See Appendix. A)~\cite{Zhongee}. This makes the driven dynamics can be accurately described by a finitely entangled state, such as the MPS in this study.

\section{The KZM and The FTS}\label{KZMFTS}
The KZM is a mechanism characterizing the production of the topological defects when a system is driven across a critical point. This mechanism was first proposed by Kibble in cosmology~\cite{Kibble}, and then in condensed matter physics by Zurek~\cite{Zurek}. Recently, the KZM has been generalized to the quantum critical dynamics~\cite{qkz1,qkz2,qkz3,qkz4,qkz5,qkz6,qkz7,qkz8,qkz9,NDAntunes,BDamski,Francuz,Chandran,Gerster}. By comparing the response time scale and the driven time scale, the KZM separates the whole process into three stages: an impulse stage sandwiched by two adiabatic stages. In the adiabatic stage, the driven time scale $\zeta_d$ is much larger than the response time scale $\zeta_r\sim \Delta^{-1}$ with $\Delta$ being the energy gap; while in the impulse stage, the response time scale is larger than the driven time scale. The KZM predicts that topological defects emerge after the quench and the number of topological defects is propositional to $R^{1/r}$, in which $R$ is the driving rate and $r$ is its dimension. The KZM has been verified numerically and experimentally in various systems, including classical and quantum phase transitions~\cite{Zurek,Kibble,qkz1,qkz2,qkz3,qkz4,qkz5}. However, the KZM assumes that the system in the impulse region ceases to evolve. It has been shown that this is an over-simplified assumption, since the system still evolves in the impulse region~\cite{Zhong1,Zhong2,Yin1,qkz6,qkz7,qkz8,qkz9,Chandran}.

The FTS focuses on the driven dynamics in the impulse region~\cite{Zhong1,Zhong2,Yin1}. One can compare it with the finite-size scaling in the space domain. The finite-size scaling shows that the lattice size $L$ characterizes the scaling properties of the macroscopic quantities when the correlation length $\xi \gg L$, namely, $|g|\ll L^{-1/\nu}$. Similarly, in the time domain, the FTS theory shows that the external driven time scale $\zeta_d\sim R^{-z/r}$ dominates the evolution of the system in the impulse region, in which $\zeta_d\ll \zeta_r$.

Based on the FTS theory for changing the transverse field~\cite{Zhong1,Zhong2,Yin1}, $f=Rt+f_i$, in which $f_i$ is far away from the critical point, the order parameter $m$ satisfies
\begin{equation}
    m(g,R) = R^{\beta/\nu r} f_1 (g R^{-1/ \nu r}),
    \label{Eq_g}
\end{equation}
where $r =1/\nu+z$ and $f_i$ is irrelevant. This is the case considered in the experiment~\cite{Keesling}, in which the scaling behavior of the correlation function is considered. Similar full scaling forms have been obtained from other arguments~\cite{qkz6,qkz7,qkz8,qkz9,Chandran}. Similarly, for the case of changing the longitudinal field according to $h=R_h t+h_i$ with $h_i$ being far away from zero, the scaling form of $m$ is
\begin{equation}
    m(g, R_h, h) = R_h^{ \beta / \nu r_h} f_2 (g R_h^{-1/\nu r_h}, h R_h^{-\beta \delta / \nu r_h} ),
    \label{Eq_h}
\end{equation}
where $r_h =\beta\delta/\nu + z$ and $h_i$ is also irrelevant.

Since in real experiments thermal effects must be involved, we also consider the case including the finite-temperature effects~\cite{Weiss,Grandi,Grandi2,Deng,Yin2}. It has been shown that~\cite{Yin2}, for the closed system, thermal fluctuations only affect the driven dynamics when the initial parameter is in the vicinity of the critical system, while for the initial parameter far away from the critical point, where $\Delta\gg T$, thermal fluctuations play negligible roles. Therefore, the thermal effects and the initial parameter should be considered simultaneously. Accordingly, the order parameter $m$ for $g=Rt+g_i$ is~\cite{Yin2}
\begin{equation}
    m(R, g_i, g, T) = R^{\beta / \nu r} f_3(g_i R^{-1/\nu r}, g R^{-1/ \nu r}, T R^{ - z/ r}),
    \label{Eq_g_T}
\end{equation}
in which $g_i$ is small and $T$ is the initial temperature. Similarly, for $h=R_h t+h_i$ and $g=0$, the order parameter $m$ satisfies
\begin{eqnarray}
    &&m(h, R_h, h_i, T) = R_h^{\beta / \nu r_h} \notag \\
    &&f_4(h_{i} R_h^{-\beta \delta /\nu r_h}, h R_h^{-\beta \delta /\nu r_h}, T R_h^{-z/ r_h}),
    \label{Eq_h_T}
\end{eqnarray}
in which $h_i$ is small.

\section{Results} \label{sec_t_0}

\subsection{FTS with changing the transverse field} \label{sec:g_0}
We first utilize the FTS to study the driven critical dynamics in the QCCM~(\ref{Eq_model}) by changing its transverse field $f$ with an adiabatic initial stage. According to Eq.~(\ref{Eq_g}), when the order parameters $m$ for different driving rates $R$ equal zero, the corresponding transverse fields $f_0$ should satisfy
\begin{equation}
    f_0(R) = f_c + cR^{1/\nu r}.
    \label{detcp}
\end{equation}
By fitting $f_0$ versus $R$ according to Eq.~(\ref{detcp}), one can obtain the critical point $f_c$ and the critical exponent $1/\nu r$. Then according to Eq.~(\ref{Eq_g}), one finds that at the critical point the order parameters $m_0$ obey
\begin{equation}
    m_0(R) \propto R^\frac{\beta}{\nu r}.
    \label{detcp1}
\end{equation}
From Eqs.~(\ref{detcp}) and~(\ref{detcp1}), one can determine the critical exponent $\beta$. Then by substituting these exponents into Eq.~(\ref{Eq_g}), one can self-consistently examine the FTS in model~(\ref{Eq_model}) and its critical properties.

\begin{figure}[tbp]
\includegraphics[angle=0,scale=0.17]{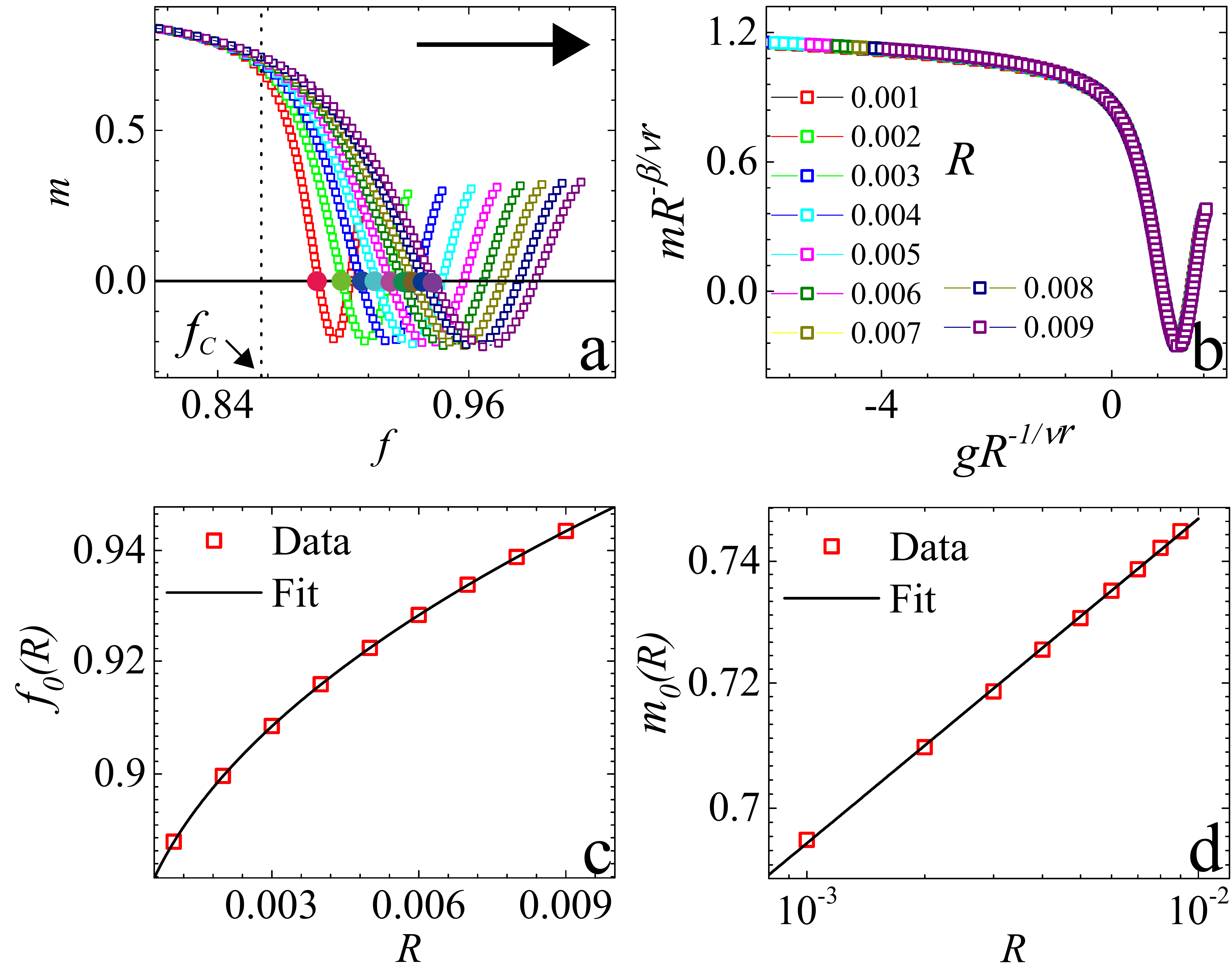}
  \caption{The evolution of $m$ under changing the transverse field {\bf a} before and {\bf b} after rescaling. {\bf c} Estimation of $f_c$ and $1/\nu r$ by by fitting $f_0$ (coloured dots in {\bf a}) versus $R$. The arrow indicates the direction of changing $f$. {\bf d} Estimation of $\beta/\nu r$ by power fitting of $m_0$ (intersection points of the dotted line and the curve of $m$ versus $f$ in {\bf a}) versus $R$.}
\label{fig:fRa}
\end{figure}

Figure~\ref{fig:fRa} shows the results for changing $f$ with different $R$. In Fig.~\ref{fig:fRa}{\bf c}, the critical point is estimated to be $f_c=0.8609$, close to $0.8612$ determined in the previous study~\cite{Samajdar}. In Figs.~\ref{fig:fRa}{\bf c} and {\bf d}, the exponents $1/\nu r$ and $\beta/\nu r$ are then estimated to be $1/\nu r=0.5039$ and $\beta/\nu r=0.03171$, respectively. By substituting these results into Eq.~(\ref{Eq_g}), we find in Fig.~\ref{fig:fRa}{\bf b} that the curves of rescaled $m$ versus $f$ collapse onto each other nicely. This result not only demonstrates that the FTS form of Eq.~(\ref{Eq_g}) is applicable in model~(\ref{Eq_model}), but also shows that the critical exponents determined from Eq.~(\ref{Eq_g}) are scientifically sound.

\subsection{FTS with changing the longitudinal field} \label{sec:h_0}
Then we study the driven dynamics for changing the longitudinal field in model~(\ref{Eq_model}). To do this, we add a symmetry-breaking term $-h\sum_i \sigma_i + \rm{h.c.}$ in Eq.~(\ref{Eq_model}). The initial value $h_i$ is very large and it is irrelevant. Similar to the case of changing $f$, we denote the longitudinal field at $m=0$ and $g=0$ as $h_0$. According to Eq.~(\ref{Eq_h}), $h_0$ for different $R_h$ satisfies
\begin{equation}
    h_0(R) \propto R_h^\frac{\beta \delta}{\nu r_h}.
    \label{detcp3}
\end{equation}
Also, at $h=0$ and $g=0$, the order parameter $m_0$ for different $R_h$ obeys
\begin{equation}
    m_0(R) \propto R_h^\frac{\beta}{\nu r_h}.
    \label{detcp4}
\end{equation}
From Eqs.~(\ref{detcp3}) and (\ref{detcp4}), one can determine the critical exponents $\beta \delta/\nu r_h$ and $\beta/\nu r_h$.

\begin{figure}[t]
\includegraphics[angle=0,scale=0.165]{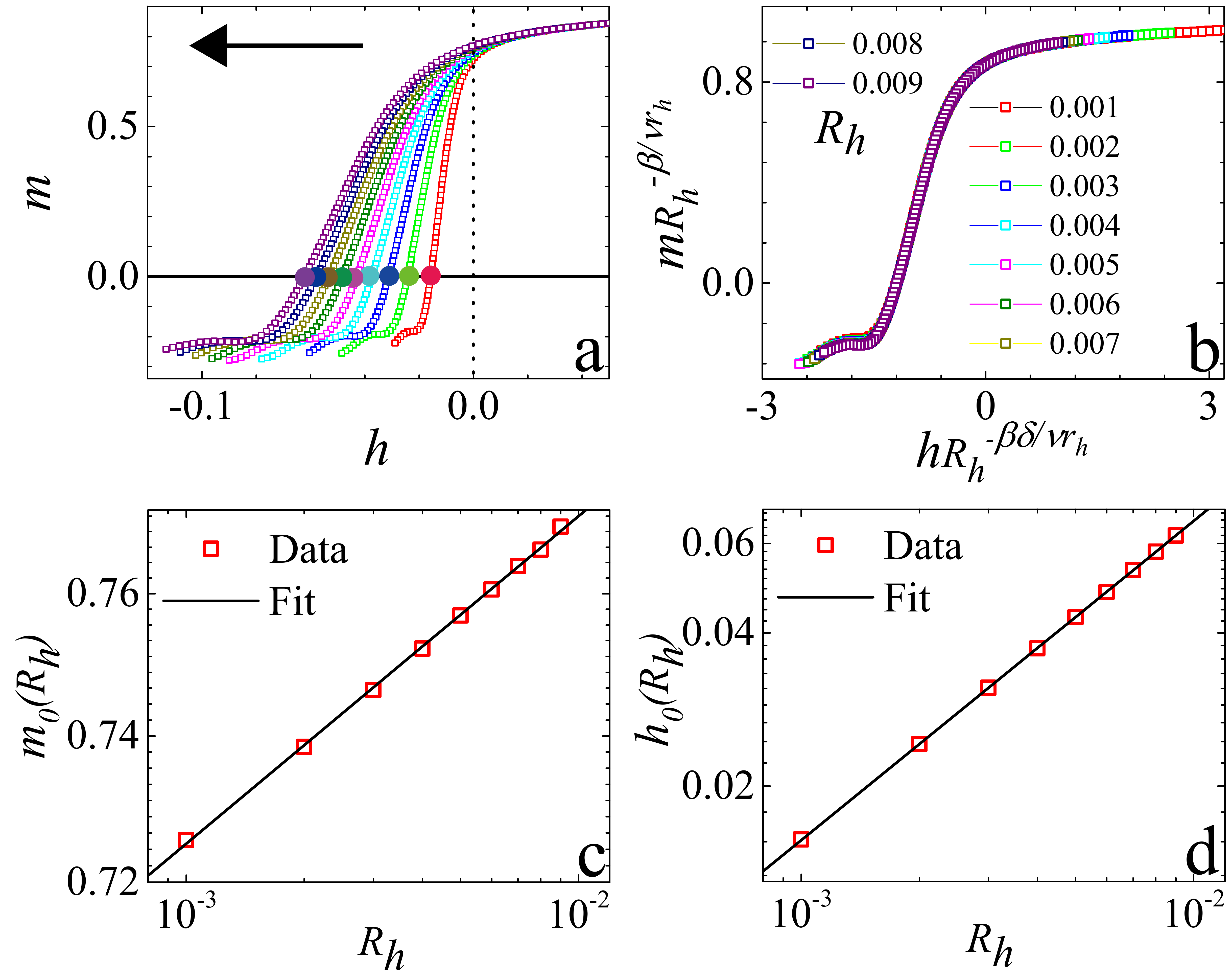}
  \caption{The evolution of $m$ under changing the longitudinal field $h$ {\bf a} before and {\bf b} after rescaling. The arrow in {\bf b} indicates the direction of changing $h$. {\bf c} Power function fitting of $m_0$ (coloured dots in {\bf a}) versus $R_h$. {\bf d} Power function fitting of $h_0$ (intersection points of the dotted line and the curve of $m$ versus $h$ in {\bf a}) versus $R_h$.}
\label{fig:hRAll}
\end{figure}

Figure~\ref{fig:hRAll} shows the results for changing $h$ with different $R_h$ at the critical point $f_c$ determined above. As shown in Figs.~\ref{fig:hRAll} {\bf c} and {\bf d}, the critical exponents $\beta \delta/\nu r_h$ and $\beta/\nu r_h$ are determined to be $\beta \delta/\nu r_h= 0.6280$ and $\beta/\nu r_h=0.0275$, respectively. Then, by substituting these exponents into Eq.~(\ref{Eq_h}), we rescale the curves of $m$ versus $h$ with $R_h$ and find that these curves collapse onto each other. This result confirms the FTS form Eq.~(\ref{Eq_h}) for model~(\ref{Eq_model}), and also shows that the exponents determined from the FTS theory are reliable.

\subsection{Table of critical exponents}
\begin{figure*}[t]
\includegraphics[angle=0,scale=0.2]{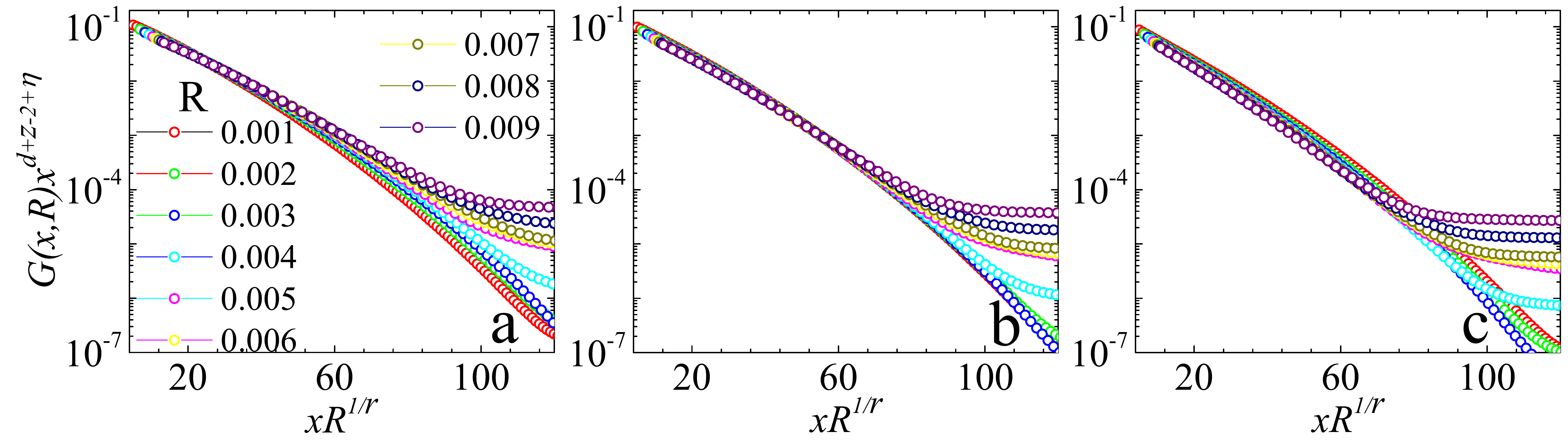}
  \caption{Attempted data collapse with different tentative values of $z$. The rescaled correlation functions collapse onto a single curve when $z=1.22$ in {\bf b} and separate from each other when slightly getting smaller at $z=1.14$ in {\bf a} and larger at $z=1.30$ in {\bf c}.}
\label{fig:cor_fun}
\end{figure*}

To the best of our knowledge, critical exponents $\beta$ and $\delta$ for model~(\ref{Eq_model}) have not been determined before. From Sec.~\ref{sec:g_0} and \ref{sec:h_0}, we can determine and verify them according to the FTS theory. In this part, we show that other exponents can also be determined independently from the FTS theory. To do this, we calculate the dynamics of correlation function $G(x)\equiv\left| \langle\sigma_{i}\sigma_{i+x}\rangle-\langle\sigma_{i}\rangle\langle\sigma_{i+x}\rangle\right|$. For changing $f$, $G$ satisfies
\begin{equation}
G(x,R) = \frac{1}{x^{d+z-2+\eta}} g(x R^{1/r}),
\label{Eq_corfun}
\end{equation}
at the critical point $f_c$. According to Eq.~(\ref{Eq_corfun}) and the scaling law $\eta=2-\beta(\delta-1)/\nu$, one can determine $z$ and $\nu$ by setting $\beta$, $\delta$, and $\nu r$ as input. Figure~\ref{fig:cor_fun} shows that the rescaled curves of $G$ versus $x$ collapse best for $z=1.22$. Accordingly, we determine all the critical exponents independently according to the FTS theory. We list the results including cases for other $\theta$ in Table.~\ref{table_exp}.

\bgroup
\begin{widetext}
\centering
\def\arraystretch{1.4}
\begin{table*}[t]
\centering
{
\bgroup
\setlength{\tabcolsep}{4.3pt}
\begin{ruledtabular}
\begin{tabular}{l l l l l l l l l l l l}
\multicolumn{1}{c}{$\theta$} &\multicolumn{1}{c}{$f_c$} &\multicolumn{1}{c}{$f_c^a$} &\multicolumn{1}{c}{$\beta$} &\multicolumn{1}{c}{$\delta$} &\multicolumn{1}{c}{$z$} &\multicolumn{1}{c}{$z^a$}  &\multicolumn{1}{c}{$\nu$} &\multicolumn{1}{c}{$\nu^a$} &\multicolumn{1}{c}{$\beta (\delta+1)$} &\multicolumn{1}{c}{$(d+z)\nu$} &\multicolumn{1}{c}{ $(d+z^a)\nu^a$}\\
\hline
$\pi/48$  & $0.9958(3)$ & $0.9960$ & $0.105(3)$ &$15.0(5)$ & $1.03(2)$ & $1.00(7)$ & $0.84(1) $ & $0.83(6) $ & $1.68(10) $ & $1.70(4) $ & $1.66(18) $\\
$\pi/12$  & $0.9383(3)$ & $0.9387$ & $0.086(4)$ &$17.7(6)$ & $1.10(4)$ & $1.07(6)$ & $0.83(3) $ & $0.80(1) $ & $1.61(13) $ & $1.74(10)$& $1.66(7)$\\
$\pi/8$  & $0.8609(2)$ & $0.8612$ & $0.063(3)$ & $22.8(10)$ & $1.22(8)$ & $1.22(7)$ & $0.81(5) $ & $0.77(2) $ & $ 1.50(13) $& $1.80(18)$& $1.71(10) $\\
$7\pi/48$ &$0.8102(4)$ &$0.8100$ & $0.050(4)$ &$28.3(15)$ & $1.30(9)$  & $1.36(6)$ &$0.80(6) $ & $0.72(1) $ & $ 1.47(19) $& $1.84(21) $ & $1.70(7) $\\
\end{tabular}
\end{ruledtabular}
}
\egroup
\caption{\label{tableexp}Critical exponents for the QCCM~(\ref{Eq_model}). The fitting error $\Delta_{f_c}$ of $f_c$ is used to determine the estimation error of the critical exponents. We fit data at $f_c\pm \Delta_{f_c}$ and use the largest difference in each exponent as the estimation error. Superscript $^a$ indicates the results from Ref.~\cite{Samajdar}. Note that the energy unit chosen in Ref.~\cite{Samajdar} is $f+J$.}
\label{table_exp}
\end{table*}
\end{widetext}
\egroup

From Table~\ref{table_exp}, one finds that values of $\nu$ and $z$ are consistent with previous results~\cite{Samajdar}. Values of $\beta$ and $\delta$ are also verified since they are involved in the calculation of $\nu$ and $z$. We confirm that the dynamic exponent $z$ is a non-interger value and increases as $\theta$ increases. In addition, we examine the hyperscaling law $\beta (1+\delta) = (d+z)\nu$ as shown in Table.~\ref{table_exp}. Although the hyperscaling law seems still right within the error bar, the expectation values for $\beta (1+\delta)$ and $(d+z)\nu$ are different, and change according to different trends as $\theta$ increases. This may indicate a possible violation of the hyperscaling law when approaching the incommensurate phase in the critical line. However, due to the relative large degree of uncertainty, more careful study needs to carry to verify the hyperscaling law.

\section{FTS at finite temperatures} \label{sec_t}
In this section, we discuss the thermal effects in driven dynamics of model~(\ref{Eq_model}). From the experimental point of view, the thermal effects cannot be completely excluded; while from the theoretical point of view, the dimension of the temperature $T$ is $z$. With a nontrivial value of $z$, whether the scaling theories proposed before are still applicable should be examined.

As discussed in Sec.~\ref{KZMFTS}, the thermal effects becomes indispensable when $f_i$ ($h_i$) is close to the critical point. First we study the critical dynamics under changing the transverse field $f=f_i + Rt$. According to the Mermin-Wagner theorem, there is no order for $1$D quantum systems at finite temperatures. So, we impose a small symmetry-breaking term $-h_i \sum_i \sigma_i + \rm{h.c.}$ in the model~(\ref{Eq_model}).

\begin{figure}[t]
\includegraphics[angle=0,scale=0.17]{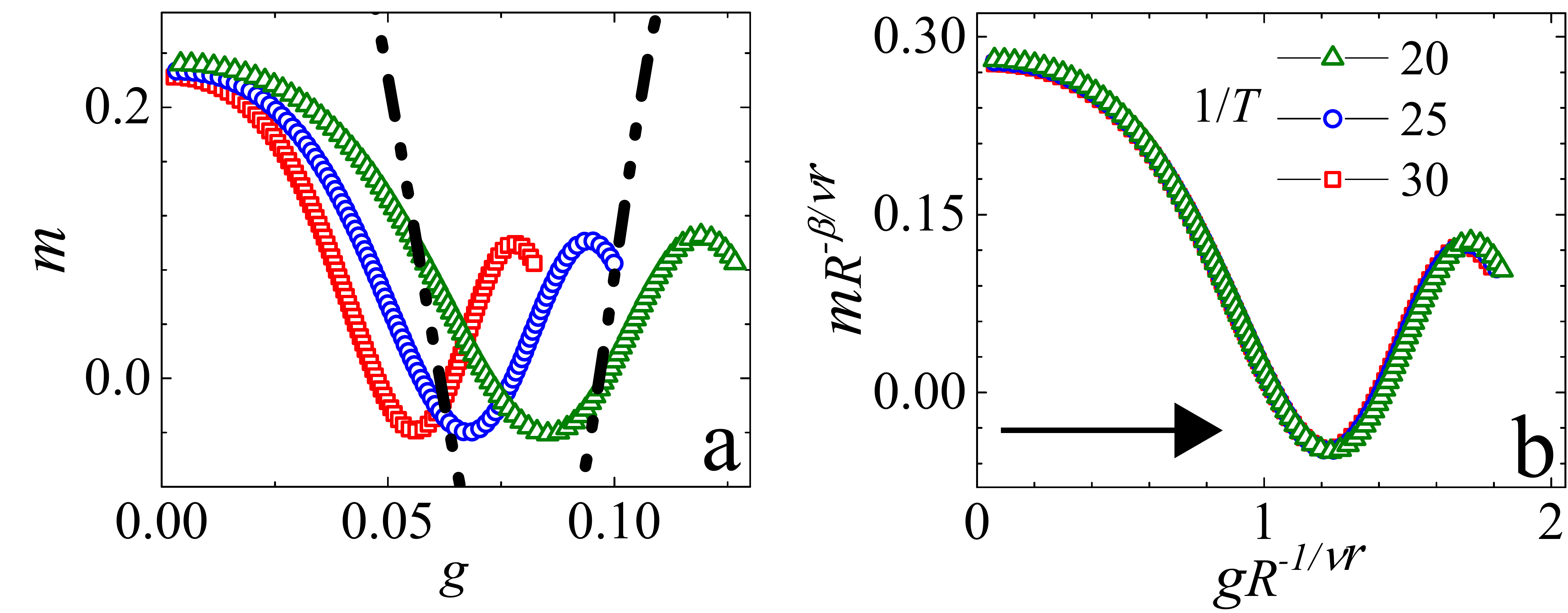}
  \caption{The evolution of $m$ {\bf a} before and {\bf b} after rescaling under changing the longitudinal field $g$ with fixed $g_i R^{-1/ \nu r} = 0.06$, $h_i R^{-\beta \delta / \nu r} = 0.08$ and $ T^{-1} R^{z/r} = 1.44$. The dash-dotted curve in {\bf a} is the evolution of $m$ starting from an adiabatic state with the same driving rate as that in the $1/T=20$ curve. The arrow in {\bf b} denotes the direction of changing $f$.}
\label{fig:gTa}
\end{figure}

We show the evolution of $m$ versus $g$ with fixed $g_i R^{-1/ \nu r}, h_i R^{-\beta \delta / \nu r}$ and $T^{-1} R^{z/r}$ at different temperatures in Fig.~\ref{fig:gTa}. Due to the thermal effect, the evolution of the order parameter is quite different from the case at the zero temperature even for the same driving rate. In spite of this, after rescaling $m$ and $g$ according to Eq.~(\ref{Eq_h_T}), we find that all the curves at different temperatures perfectly collapse onto a single one. This result confirms that the modified FTS of Eq.~(\ref{Eq_g_T}) for a non-integer $z$.

\begin{figure}[t]
\includegraphics[angle=0,scale=0.17]{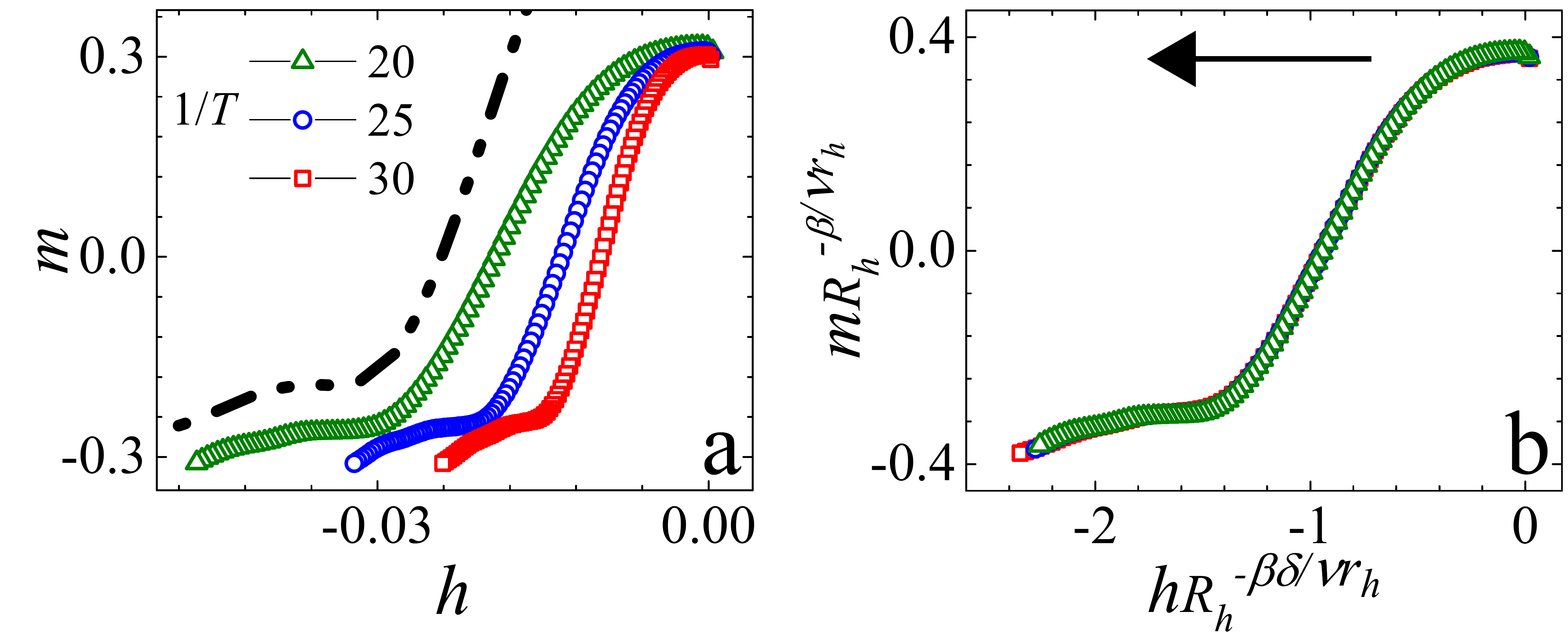}
  \caption{The evolution of $m$ {\bf a} before and {\bf b} after rescaling under changing the longitudinal field $h$ with fixed $h_i R_h^{-\beta \delta / \nu r_h} = 0.015$, $T R_h^{-z / r_h} = 0.52$ at the critical point. The dash-dotted curve in {\bf a} is the evolution of $m$ starting from an adiabatic state with the same driving rate as that in the $1/T=20$ curve. The arrow in {\bf b} denotes the direction of changing $h$.}
\label{fig:hTa}
\end{figure}

Similarly, we also study the scaling behavior including the thermal effects under changing the longitudinal field. For simplicity we consider the case in which the initial $f$ is exactly at the quantum critical point $f_c$. Figure~\ref{fig:hTa} shows that the curves of $m$ versus $h$ for different temperatures with fixed $h_i R_h^{-\beta \delta / \nu r_h}$, $T R_h^{-z / r_h}$. Althgouth thermal effects make $m$ evolve quite differently compared to the case at zero temperature even for the same driving rate, all the rescaled curves collapse onto a single one after rescaling, confirming the FTS of Eq.~(\ref{Eq_h_T}).

\section{Summary}\label{sec_summary}
We have studied the driven critical dynamics of the $1$D $\mathbb{Z}_3$ QCCM. Both cases of changing the transverse field and changing the longitudinal field have been considered. The FTS has been confirmed for a nontrivial non-integer dynamic exponent $z$. From the FTS scaling form, the critical point and the critical exponents of $\mathbb{Z}_3$ QCCM have been determined and verified self-consistently. In particular, to the best of our knowledge, critical exponents $\beta$ and $\delta$ have been estimated for the first time. Besides, the critical exponents $z$ and $\nu$ have also determined independently, and these results are consistent with previous studies. From the nonequilibrium aspect, we have confirmed that $z$ is non-integer and increases as $\theta$ increases. In addition, the thermal effects in the driven dynamics have also been studied. Our present study poses a question that whether the hyperscaling law is violated for larger $\theta$. The properties explored in the present work could be examined in future experiments.

\section*{Acknowledgments}
We thank T. Xiang and F.-C. Zhang for helpful discussions. This work is supported by the Strategic Priority Research Program of Chinese Academy of Sciences (Grant XDB28000000), and the National Science Foundation of China (Grant NSFC-11674278). S. Y. is supported in part by China Postdoctoral Science Foundation (No. 2017M620035).

\appendix
\section{The convergence of the numerical results in driven critical dynamics}
\begin{figure}[tbp]
\includegraphics[angle=0,scale=0.17]{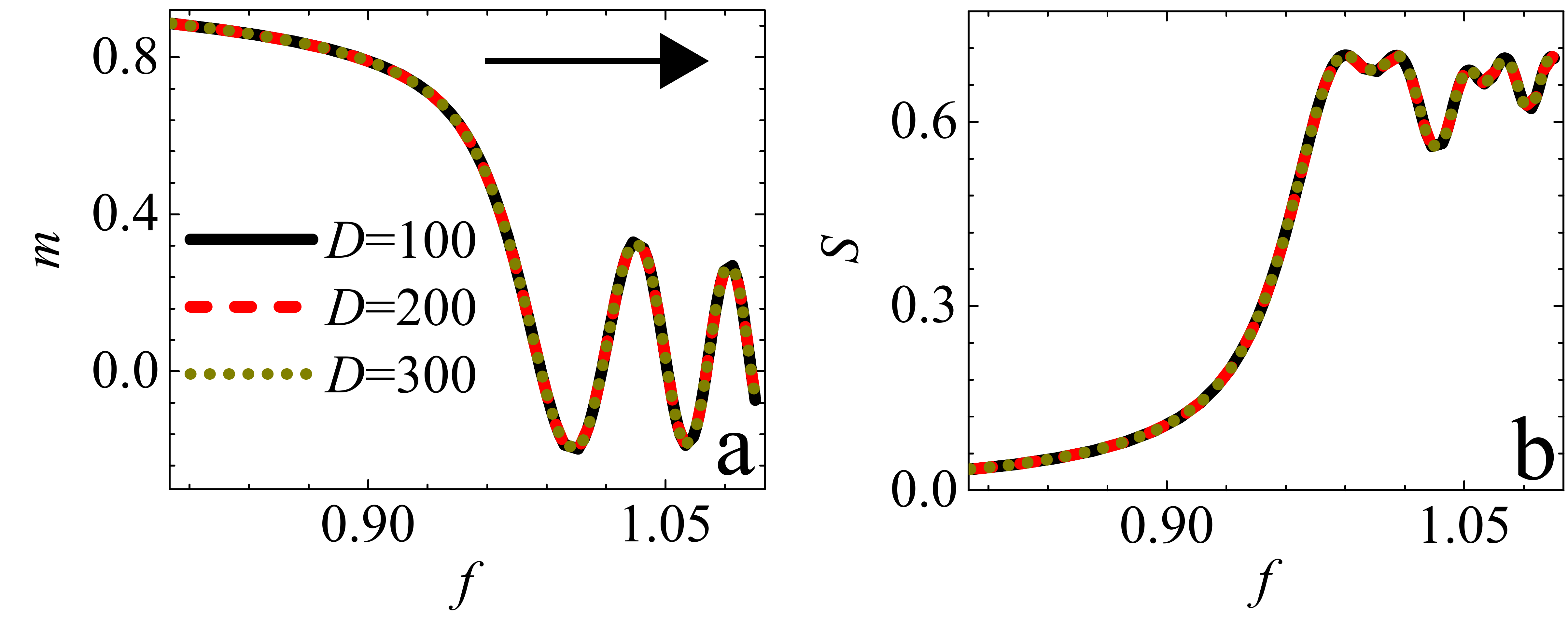}
  \caption{The evolution of the order parameter $m$ and the entanglement entropy $S$ calculated using MPS with different bond dimension $D$ are shown in {\bf a} and {\bf b} respectively for the QCCM at $\theta = \pi/12$ under changing the transverse field with the driving rate $R=0.004$.}
\label{fig:finite_D}
\end{figure}

In general, the entanglement entropy of the ground state of a $1$D critical system diverges. On the contrary, in the driven dynamics, it has been shown that the entanglement entropy is bounded due to the presence of an external driving field~\cite{Zhongee}. So the driven critical dynamics can be accurately simulated by the MPS method even with a moderate $D$. This is confirmed in Fig.~\ref{fig:finite_D}. With different bond dimensions $D$, no apparent discrepancy is found for both the order parameter $m$ and the entanglement entropy $S$. This confirms the convergence of our results and the reliability of our calculation.

\end{CJK*}

\end{document}